\documentclass[]{interact}

\usepackage{epstopdf}% To incorporate .eps illustrations using PDFLaTeX, etc.
\usepackage[caption=false]{subfig}% Support for small, `sub' figures and tables

\usepackage[longnamesfirst,sort]{natbib}% Citation support using natbib.sty
\bibpunct[, ]{(}{)}{;}{a}{,}{,}% Citation support using natbib.sty
\renewcommand\bibfont{\fontsize{10}{12}\selectfont}% To set the list of references in 10 point font using natbib.sty

\theoremstyle{plain}% Theorem-like structures provided by amsthm.sty

\theoremstyle{definition}

\theoremstyle{remark}

\begin{document}

%\articletype{ARTICLE TEMPLATE}% Specify the article type or omit as appropriate

\title{ChatGPT in the classroom. \\
Exploring its potential and limitations in a Functional Programming course. 
}

\author{
\name{M. Popovici\textsuperscript{a}\thanks{CONTACT M. Popovici. Author. Email: matei.popovici@upb.ro}}
\affil{\textsuperscript{a}POLITEHNICA University of Bucharest, Computer Science dDepartment. Splaiul Independentei 313, Bucharest, Romania.}
}

\maketitle

\begin{abstract}
In November 2022, OpenAI has introduced ChatGPT – a chatbot based on supervised and reinforcement learning. Not only can it answer questions emulating human-like responses, but it can also generate code from scratch or complete coding templates provided by the user. ChatGPT can generate unique responses which render any traditional anti-plagiarism tool useless. Its release has ignited a heated debate about its usage in academia, especially by students. We have found, to our surprise, that our students at POLITEHNICA University of Bucharest (UPB) have been using generative AI tools (ChatGPT and its predecessors) for solving homework, for at least 6 months. We therefore set out to explore the capabilities of ChatGPT and assess its value for educational purposes. We used ChatGPT to solve all our coding assignments for the semester from our UPB Functional Programming course. We discovered that, although ChatGPT provides correct answers in 68\% of the cases, only around half of those are legible solutions which can benefit students in some form. On the other hand, ChatGPT has a very good ability to perform code review on student programming homework. Based on these findings, we discuss the pros and cons of ChatGPT in a teaching environment, as well as means for integrating GPT models for generating code reviews, in order to improve the code-writing skills of students.
\end{abstract}

\begin{keywords}
learning and generative AI, Empirical Studies of Programming and Software Engineering. Human Language Tehnologies in program development.
\end{keywords}

\section{Introduction}

ChatGPT has experienced a surge in popularity in recent times. There is a great deal of lively debate surrounding its potential uses, impact on research: \cite{StokelWalker2023}, science: \cite{Frieder2023}, and education: \cite{Lambeets2023, Elsen-Rooney2023, Jalil2023, Lau2023}. Since its increase in popularity in November 2022, academia has been testing its limits with surprising results in its apparent proficiency to address a wide range of questions. ChatGPT is the most recent addition to a series of generative pre-trained transformers (GPTs) that utilize language AI models trained on vast amounts of human text. Its purpose is to generate discourses that closely resemble those produced by humans. In brief, ChatGPT relies on word probabilities to predict the most meaningful continuation of a current sentence. In addition, ChatGPT incorporates a degree of randomness, resulting in text that appears more natural and human-like. Due to its ability to generate coherent discourse, ChatGPT is an attractive option for students who wish to utilize it for various homework assignments: written essays, argumentation, coding. Additionally, its ability to generate unique text renders all traditional plagiarism tools ineffective, adding to its appeal as a valuable resource for students. 

While the debate about ethics and long-term implications of ChatGPT carries on, there is also a separate debate, among students, about the role of ChatGPT in learning and access to correct information. We have conducted a survey among more than 180 students which shows that at least 40\% of respondents have used generative AI tools for solving homework on at least one occasion. At the same time, only 28\% of students believe that this has benefited them in actual learning or in a better understanding of the assignment. It is well-known that ChatGPT is not designed to provide references for the text it generates, thus making it hard, if not impossible to trace every statement to assess its validity. Despite that, at our University reports have surfaced anecdotally of students favoring ChatGPT answers as a study resource over lecture notes to prepare for exams. While ChatGPT eliminates the need for reformulation and narrative discourse development, it also underscores the importance of the verification and validity of information sources.

Our work is an attempt to explore this problem, in the limited setting of an undergraduate programming course taught at POLITEHNICA University of Bucharest (UPB) to undergraduate second-year students. We are interested in exploring how and to what extent ChatGPT can benefit both students and tutors in supporting the educational process in a positive and constructive manner. We firmly believe that ChatGPT is a valuable tool and that forbidding it, by ignorance or by rules, as was the case in New York City where ChatGPT was blocked (Elsen-Rooney, 2023), is both impractical and lacking in innovation. 

Our goal is to gain a deeper understanding of the tool and its limitations, while also exploring healthy ways for its application in a programming-focused lecture. We have initiated an extensive survey to gain insights into our students' relationship with ChatGPT. We have found that: (i) over 40\% of the respondents have been using ChatGPT or similar tools for at least 6 months for coding assignments, and (ii) over 38\% of the respondents are confident with the answers provided by ChatGPT. The complete details of the survey are discussed in Section 2.

Subsequently, we studied the actual precision of ChatGPT in solving programming assignments. We subjected our complete corpus of 72 programming tasks, which covers the entire laboratory of our Functional Programming lecture (Popovici, 2022) to ChatGPT to evaluate its performance. Laboratories include mostly medium to short programming assignments which students usually solve in class with the help of a Teaching Assistant. In our study, a solution generated by ChatGPT must: (i) compile as well as (ii) pass a series of tests to be deemed correct. In addition to these two distinct criteria, we include two more: (iii) a solution must be educational in some sense - students should have the opportunity to learn from reading it, and (iv) efficient - it must use sound coding practices which ensure code legibility as well as efficiency. The latter two are our primary teaching goals in our course. We were surprised to see that ChatGPT did very well on our tasks. ChatGPT as a student would receive an approximate score of 7 out of a maximum of 10. Nonetheless, 43\% of the accurate solutions provided by ChatGPT are either inefficient or comprise of code that is incomprehensible for the average student. We were also interested to see to what extent ChatGPT is able to correct itself when errors within its answers are highlighted. We discovered that ChatGPT could enhance its score from 7 to 8.6, once a follow-up question or issue was highlighted. However, prompting such issues requires a certain level of expertise from the average student. To this end, we also used ChatGPT to generate tests for each of its solutions. We have found that only 70\% of generated tests are correct, and not even follow-up questions can improve this result. Our labs do not provide tests for students in our assignments. Thus, when a test fails, students may find it difficult to determine whether the issue lies with the solution or in the test supplied by ChatGPT. The results for our evaluation study as well as an in-depth comparison between human and code-generated errors are detailed in Section 3. 

Lastly, during our assessment, we noticed that ChatGPT can accurately respond to qualitative questions concerning the code we submitted. Examples of such questions are: “is the code functional?”, “does the code use side-effects?”, “is the given function tail-recursive?”. This observation prompted us to develop a tool which relies on ChatGPT for code-reviewing submitted homework. Unlike testing, which ensures the correctness of a submitted solution, the objective of code review is to assess and possibly improve the legibility and structure of the code. In our lecture, it is particularly important to review if the code was written in the functional style. Our team usually goes over more that 60 submissions, for each of the 4 homework that we give in the Functional Programming lecture. Although code-review is highly beneficial for students, conducting it comprehensively can be extremely time-consuming and resource-intensive for our team. Our tool relying on ChatGPT can automate most of this process, with high accuracy. Code reviewers can select dedicated parts of the homework under review (e.g. functions with specific names), then formulate questions such as the ones previously illustrated. Our tool will parse the desired code fragment from each submitted homework and combine it with the question we are targeting. Next, it will perform a sequence of ChatGPT queries, one for each homework, and retrieve each answer. These can subsequently be subject to human review before being submitted as feedback to students. We supply more details on this in Section 4.

We discuss the implications of our observations, as well as ways for mitigating the use of ChatGPT and other similar tools, in Section 5. In Section 6 we review major trends in educational tools for assessment and feedback prior to ChatGPT, and relate them to our own classroom experiences and practices. In Section 7 we review related work and in Section 8 we conclude and provide future directions.

\section{ChatGPT and the student body}\label{sec:survey}
\begin{figure}
\centering
\subfloat[]{%
\resizebox*{7cm}{!}{\includegraphics{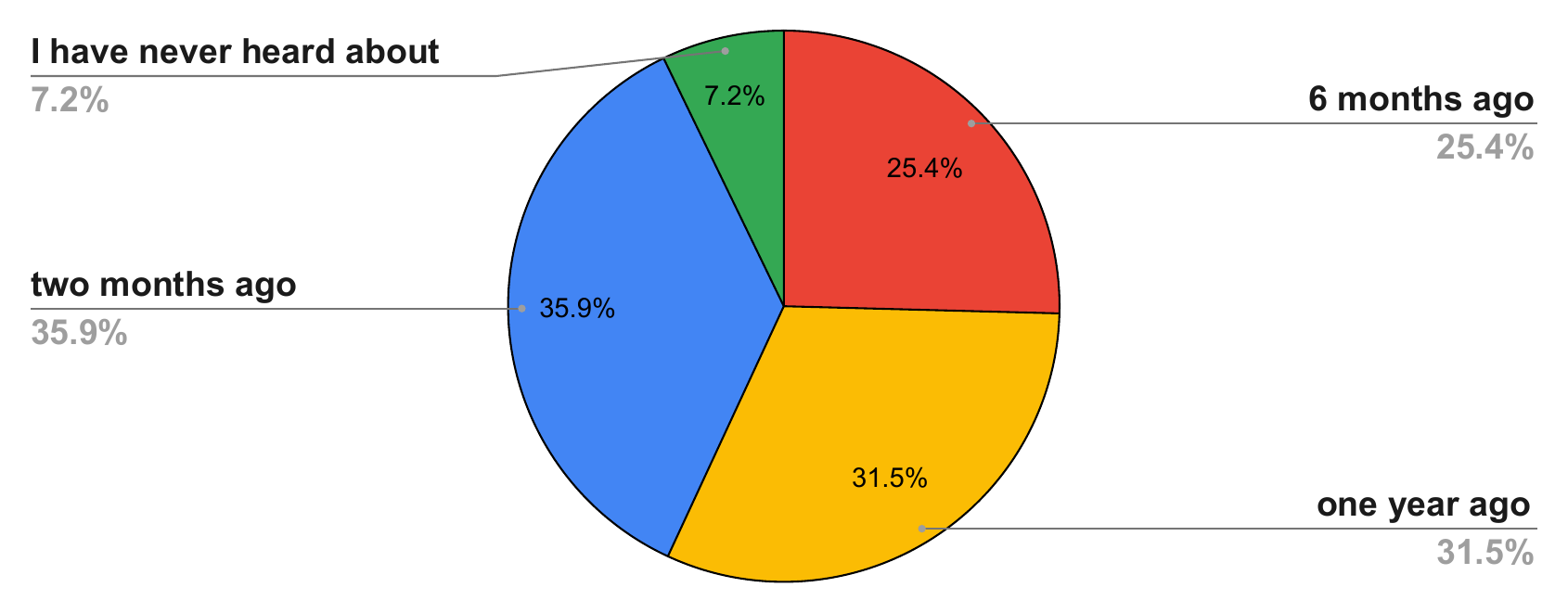}}}\hspace{5pt}
\subfloat[]{%
\resizebox*{7cm}{!}{\includegraphics{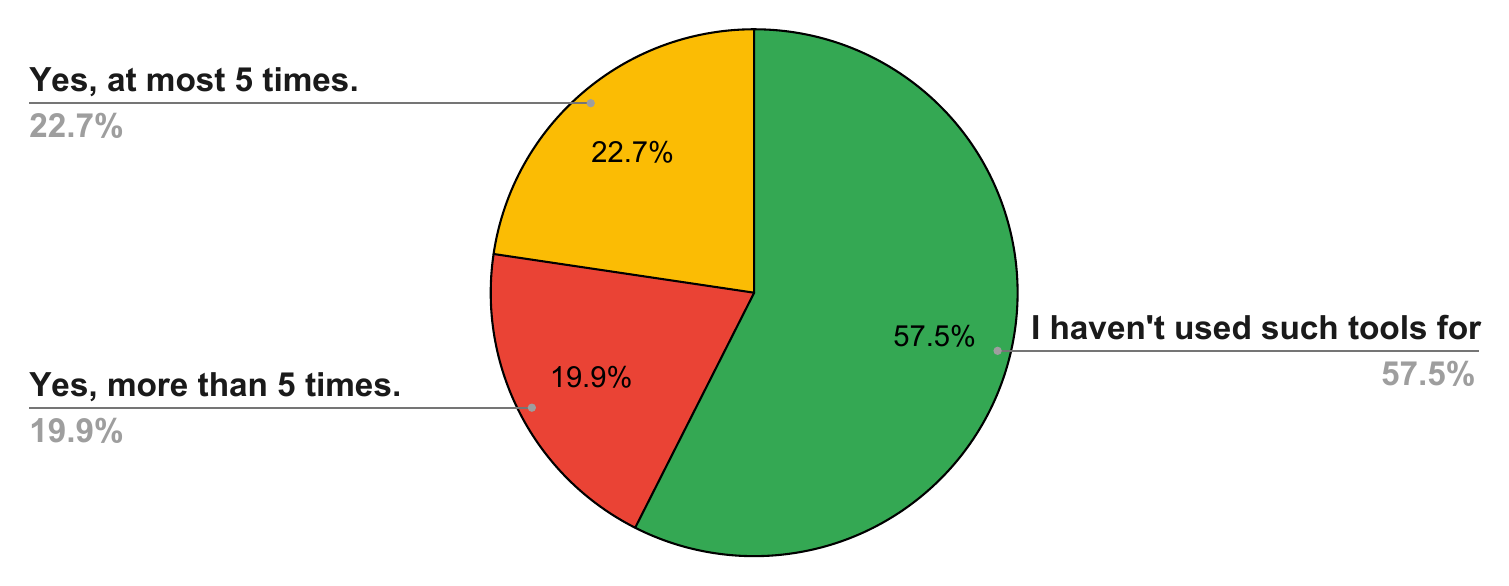}}}
\caption{Survey results: answers to questions (a) \emph{When did you hear about generative AI?} and (b) \emph{How many times have you used generative AI for homework or other school activities?}} \label{gen1}
\end{figure}

\begin{figure}
\centering
\subfloat[]{%
\resizebox*{7cm}{!}{\includegraphics{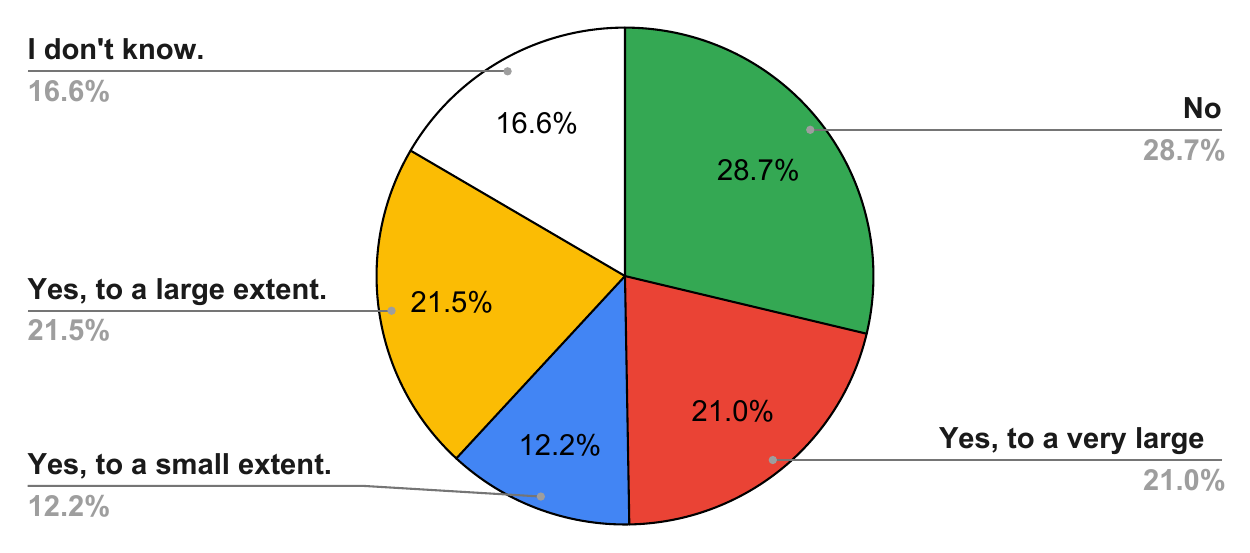}}}\hspace{5pt}
\subfloat[]{%
\resizebox*{7cm}{!}{\includegraphics{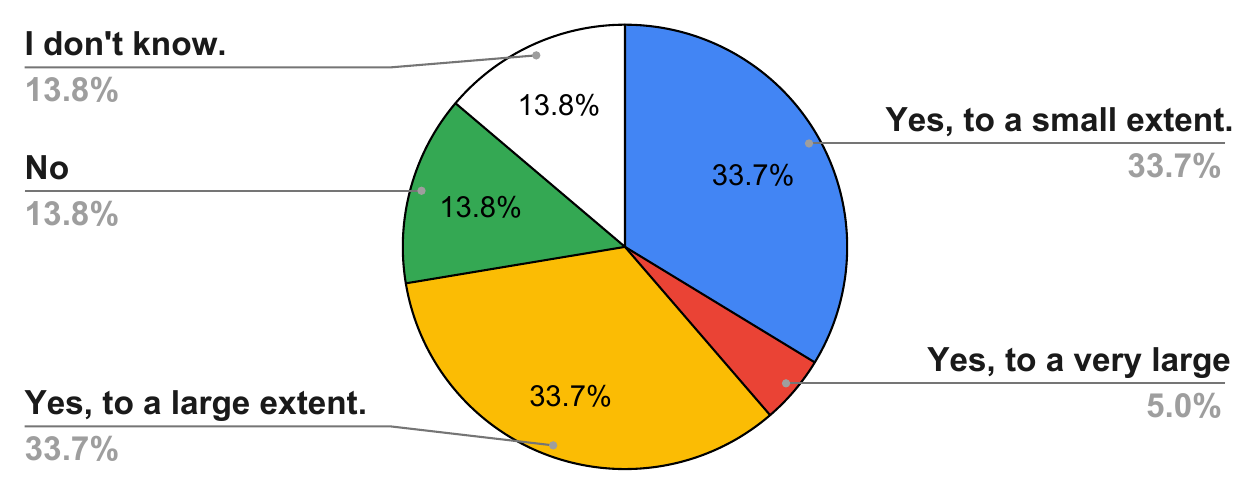}}}
\caption{Survey results: answers to questions (a) \emph{Has generative AI helped you gain a better understanding of curricula?} and (b) \emph{Do you believe generative AI has good accuracy?}} \label{gen2}
\end{figure}

At the beginning of January 2023 we have performed a survey at POLITEHNICA University of Bucharest and received 181 responses from students mostly ranging in the second and third years. Every response was entirely anonymous. Students were under no obligation to participate or respond in any particular manner. All contributions were voluntary and self-directed. The questions were not in regards to any specific lecture of group of lectures taught at our University. Instead, they encompassed their overall academic coding experience, as a whole. We have found that, within the last year, 31\% of students became aware of code generation tools, while over 55\% became aware of them within the last 6 months of the survey time (Figure~\ref{gen1} (a)). Over 40\% of responding students have reported using generative AI for homework and lab assignment (Figure~\ref{gen1} (b)). 
Roughly a third of responding students believe generative AI has good or perfect accuracy, meaning that it generates code which is correct and compiles. At the same time, a third of responders believe such tools to have low accuracy and 13\% believe they are not accurate at all (Figure~\ref{gen2} (a)). Finally, when asked whether generative AI has been helpful in solving programming assignments, 42\% of respondents believe it has been helpful, while 40\% believe it has provided little to no assistance (Figure~\ref{gen2} (b)). 

The survey clearly indicates that tools such as ChatGPT are utilized on a regular basis by students in writing homework and pursuing exams. For this reason, we would like to gain a better understanding into the accuracy and practicality of generative AI tools, in particular ChatGPT, in providing correct but also useful results for students.

\section{ChatGPT Evaluation}

\subsection{Dataset description and methodology}

\begin{figure}[h]
\centering
\subfloat[]{%
\resizebox*{4.5cm}{!}{\includegraphics{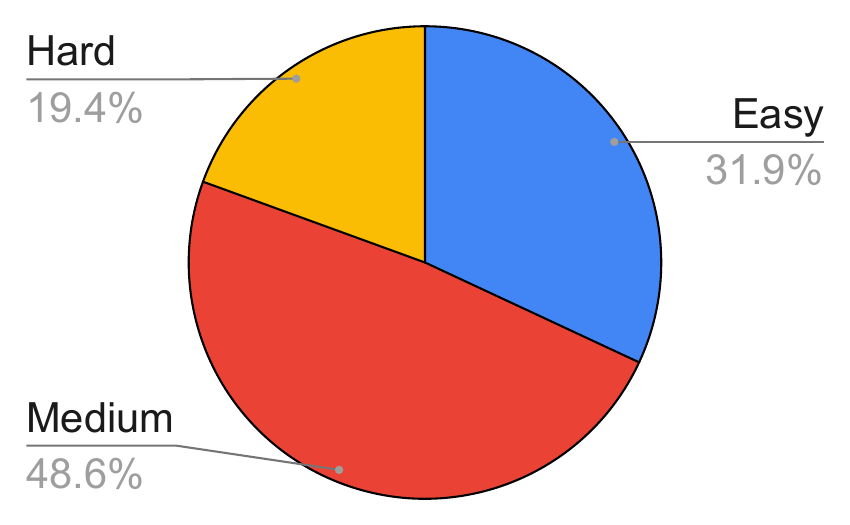}}}\hspace{5pt}
\subfloat[]{%
\resizebox*{4.5cm}{!}{\includegraphics{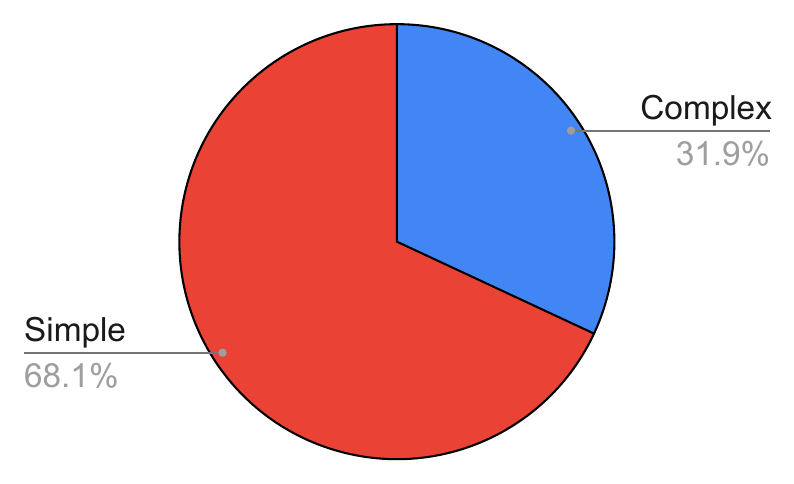}}}
\subfloat[]{%
\resizebox*{4.5cm}{!}{\includegraphics{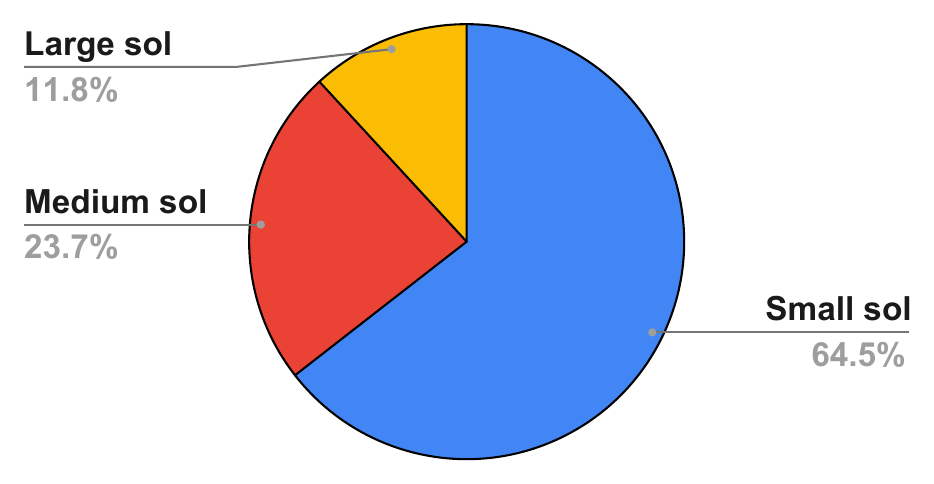}}}
\caption{Breakdown of our dataset into:  (a) Hard, medium and easy exercises (b) Exercises with simple/complex statements (c) small, medium or large-sized solutions.} \label{breakdown}
\end{figure}
Our dataset consists of the entire corpus of exercises spanning 7 labs from the lecture “Functional Programming” (FP) taught in the Scala programming language (designed by~\cite{Odersky2004}) during the second year of a Computer Science engineering degree program at UPB. It contains 72 coding exercises, which we grouped into three categories: easy (23 exercises), medium (35 exercises) and hard (14 exercises), based on our teaching experience from previous years.

We tagged each exercise as having a simple (49 exercises) or complex statement (23 exercises). Simple statements are single-phrased and may be followed by a simple function signature. Complex statements are multi-paragraph, may contain several code snippets which are relevant for the solution or alternatively may contain mathematical equations described using ASCII symbols (e.g. \texttt{x\_1 + 1 = x\_2}). Many of our exercises rely on templates, wherein the students are required to fill-out a function signature or a code structure provided to them. An example of such a template is the following:

\begin{verbatim}
def factorial (n: Int): Int = ???
\end{verbatim}

A small number of exercises (9 in number) are not template-based. Their statement is defined solely in natural language.  Finally, we have also classified each exercise according to the size of the solution, more precisely its number of lines of code (loc), into: small (1 to 6 loc) - 49 exercises, medium (6-12 loc) - 18 exercises and large (over 12 loc) - 5 exercises. The distribution of exercises based on difficulty, statement complexity and solution size is illustrated in Figure~\ref{breakdown}.

The version of ChatGPT utilized in our experiments was the one released on January 30th. To maintain context, each interaction with ChatGPT was conducted seamlessly in a single chat session, to allow ChatGPT to learn from previous answers and to preserve context. Some sets of exercises may require individual solutions which, used in combination, can be utilized to solve a more intricate problem. For instance, computing the square root of a natural number using Newton’s Approximation Method requires a function which: (i) improves by one iteration one estimation of the square root and another which (ii) checks if an estimation is accurate enough (by a given threshold). These are split along several exercises.

We compiled each solution generated by ChatGPT using the Scala compiler. We have found that most solutions compile, hence we focused on their correctness which we evaluated using our own tests. We have recorded all valid solutions from the initial ChatGPT response. If a solution was incorrect (with some tests failing or, less frequently, being unable to compile) we replied to ChatGPT that there was an issue, without any indication of the specific problem. If the second follow-up answer did not produce a correct solution, we tried to steer ChatGPT towards the root-cause of the problem. We wanted to see if ChatGPT can improve, however this usage scenario is less likely from a student’s viewpoint, since we do not provide tests for students in our lab. We find it less likely that students which rely on ChatGPT will be able to write good test-suites for their solutions.
We have found that, in many instances, if ChatGPT’s second solution was incorrect, our recommendations were not useful in enhancing its response. If the answer remained incorrect, we stopped after the second suggestion and recorded whether the answer was improved. We used ChatGPT to generate tests for each solution and followed the same procedure as previously described (a maximum of two follow-up questions if the test was incorrect). Lastly, we documented whether the final solution was legible or not. An illegible solution contains code or code-structure which is not necessarily incorrect, but it is either unnatural or difficult to understand from a learning perspective. We provide two examples (generated by ChatGPT):

\begin{verbatim}
def take(n: Int, l: List[Int]): List[Int] = {
	def go(n: Int, result: List[Int], l: List[Int]): List[Int] =
		(n, l) match {
			case (0, _) => result.reverse
			case (_, Nil) => result.reverse
			case (n, h :: tail) => go(n - 1, h :: result, tail)
			}
	go(n, Nil, l)
}
\end{verbatim}

The function \texttt{take} creates a new list containing only the first \texttt{n} elements of list \texttt{l}. The natural solution to implement \texttt{take} is using simple recursion. The solution given by ChatGPT relies on a helper function \texttt{go}, which is tail-recursive and has the property of extracting elements in reverse order. For this reason, before returning the result, the list must be reversed. This behavior is not yet understood at this stage of our lecture, and students not familiar with tail-recursion may have a hard time understanding why reversal is necessary. This type of an illegible solution (although correct and effective) will not help students in comprehending the task and may result in them incorrectly assuming that reversal is a necessary component of the take implementation.

The second example is a solution for a task related to the main diagonal of a matrix. The two nested \texttt{for}’s (over \texttt{i} and \texttt{j}) generated by ChatGPT are not necessary and are inefficient. They should be replaced by a single for loop ranging over the line/column of the matrix.

\begin{verbatim}
for (i <- 0 until n) {
 for (j <- 0 until m) {
   diagImg(i)(j) = img(i)(j)
   if (i == j || i + j == n - 1) diagImg(i)(j) = 1
 }
}
\end{verbatim}

\subsection{Evaluation results}\label{sec:evaluation}

We have found that ChatGPT performs surprisingly well and for two exercises it actually produced better code than our official solutions. However, the “first response” solution is correct only in 68\% of the cases, which is roughly two-thirds. Thus, ChatGPT’s performance is comparable to that of an average student. 

\begin{figure}[h]
\centering
\subfloat[]{%
\resizebox*{7cm}{!}{\includegraphics{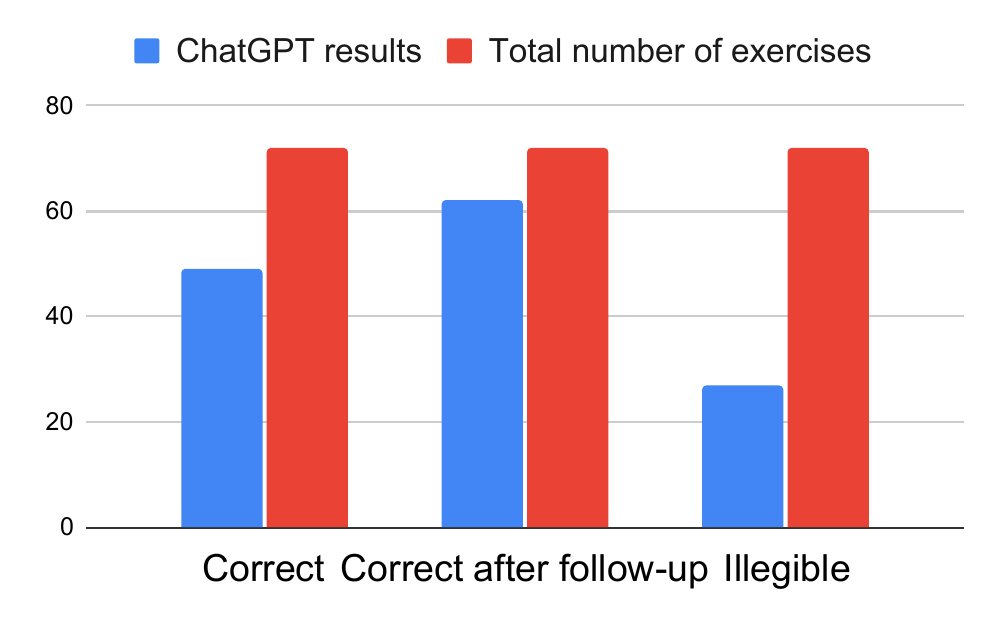}}}\hspace{5pt}
\subfloat[]{%
\resizebox*{7cm}{!}{\includegraphics{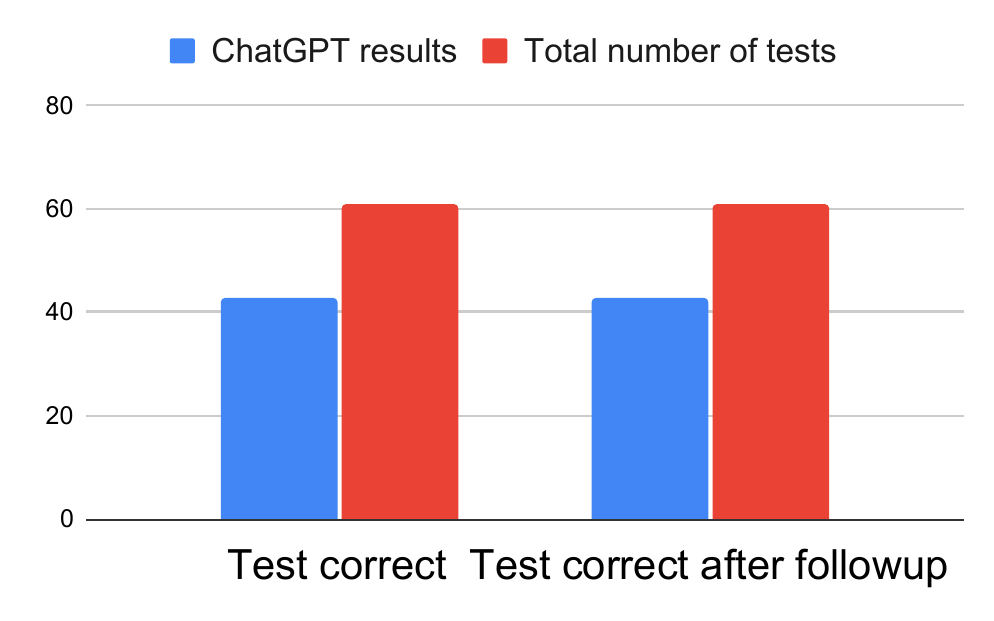}}}
\caption{Evaluation results: (a) ChatGPT exercise correctness rates and (b) correct test generation rates} \label{correctness}
\end{figure}

Once one or more follow-up questions are given, ChatGPTs improvement is substantial: 86\% of the generated solutions are correct. However, the percentage of legible solutions out of all correct ones is 57\%, which means that almost one in two correct answers may not actually be insightful for the student, either because the code is hard to understand, or because it is not efficient or properly written. In 60\% of the cases when a follow-up question was asked, the answer was helpful in the sense that ChatGPT improved on the previous solution, either by giving a better code structure which is still incorrect or by immediately identifying the correct solution. Figure~\ref{correctness} (a) illustrates our results on correctness.

\begin{figure}
\centering
\subfloat[]{%
\resizebox*{8cm}{!}{\begin{tabular}{|c|c|c|}
\hline
  & Simple statements & Complex statements \\
  Correct in the first reply & 75\% & 52\% \\
  Correct after follow-up questions & 89\% & 78\% \\ 
  \hline 
\end{tabular}}}\hspace{5pt}
\subfloat[]{%
\resizebox*{7cm}{!}{\includegraphics{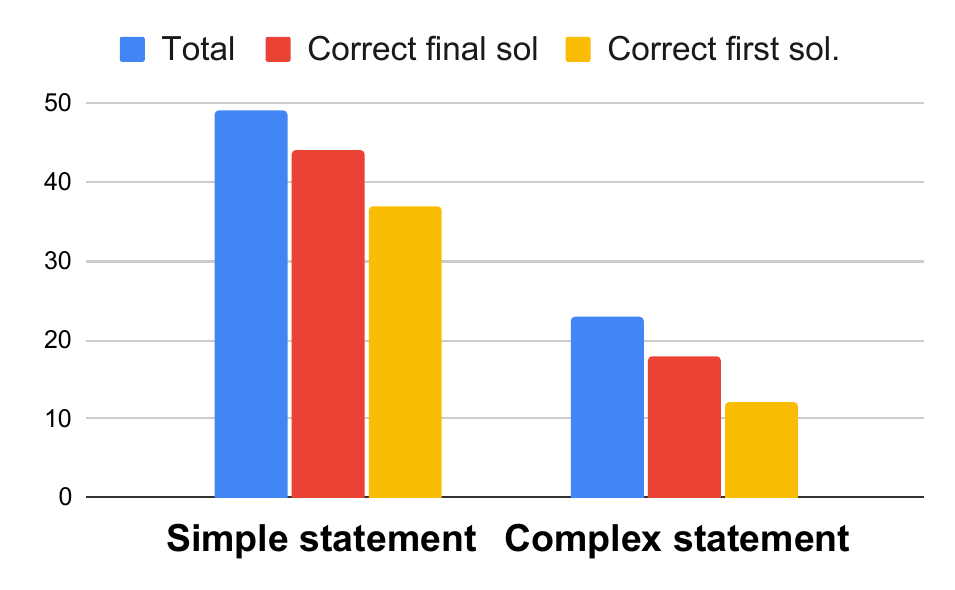}}}
\caption{Correctness of ChatGPT solutions per statement complexity, expressed as: (a) percentages (b) as absolute values from the dataset} \label{complexity}
\end{figure}

The percentage of correct tests that was generated from the first response is comparable to that of the solutions: 70\%. Unfortunately, ChatGPT is unable to improve on most of its incorrect tests. After follow-up questions the percentage of correct generated tests increases only to 72\% (Figure~\ref{correctness} (b)). ChatGPT is unable to infer basic mathematical facts. One such example is finding the divisors of 7 out of a range of integers. When asked if 7 is a divisor of 21 (7 was generating an incorrect test in one exercise), ChatGPT was unable to reply correctly. The answers to follow-up questions did not help in improving the answer.

\begin{figure}[h]
\centering
\subfloat[]{%
\resizebox*{8cm}{!}{\begin{tabular}{|c|c|c|c|}
\hline
  &Easy exercises & Medium exercises & Hard exercises \\
  Correct in the first reply & 69\% & 71\% & 57\% \\
  Correct after follow-up questions & 91\% & 88\% & 71\% \\ 
  \hline 
\end{tabular}}}\hspace{5pt}
\subfloat[]{%
\resizebox*{7cm}{!}{\includegraphics{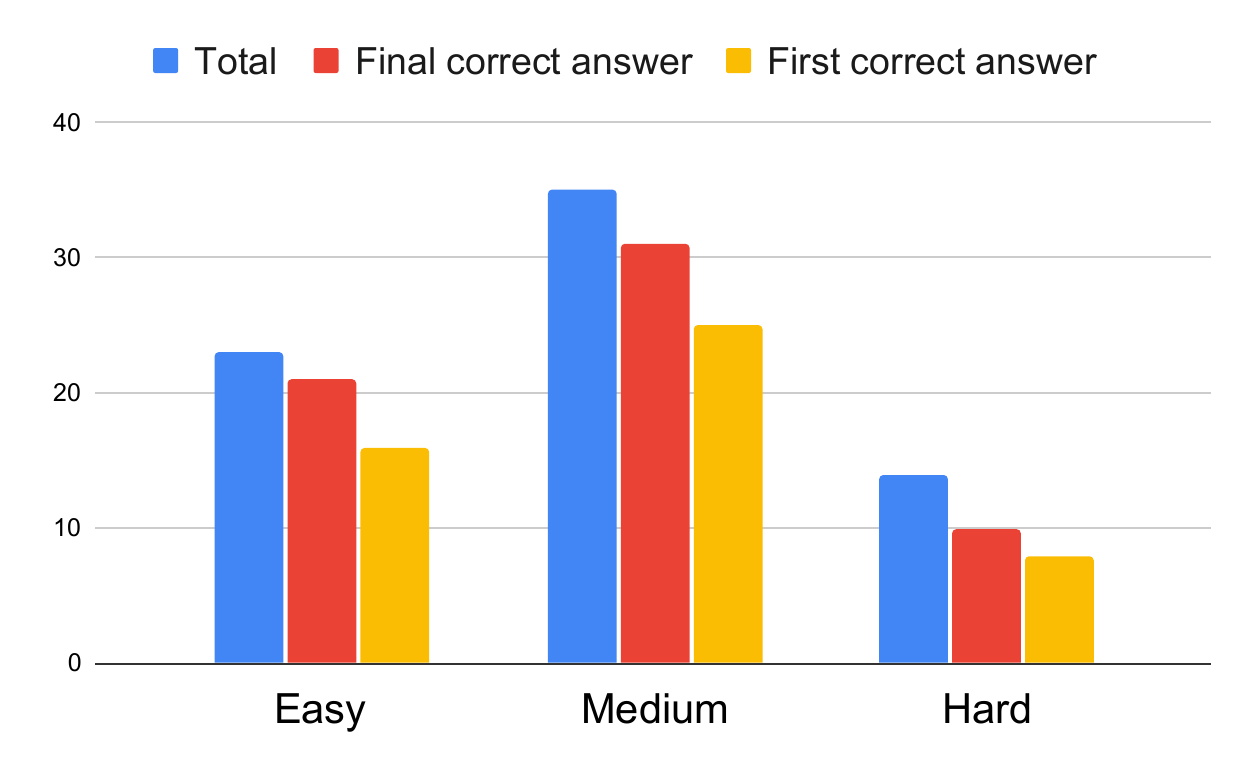}}}
\caption{Correctness of ChatGPT solutions per exercise difficulty, expressed as: (a) percentages (b) as absolute values from the dataset} \label{difficulty}
\end{figure}

We were interested to see whether ChatGPT’s correct answers correlate with features of our exercises. We discovered that, for exercises involving complex statements, the initial correct answer rate was 52\%, which increased to 78\% after follow-up questions. On the other hand, for exercises with basic statements, the correct answer rate was 75\%, improving to 89\% after follow-up questions (Figure~\ref{complexity}). Another interesting correlation to explore is that with student difficulty. ChatGPTs performance is almost indistinguishable at the easy and medium levels, however ChatGPT is sensibly less able to answer hard questions correctly (Figure~\ref{difficulty}). These questions contain subtleties in code which ChatGPT is unable to capture. One such example is related to an overflow/underflow during division in the following program statement:

\begin{verbatim}
abs(a - x*x) < threshold
\end{verbatim}

For very small values of a and x, the value of the left-hand-side of the expression may become very large, thus incorrectly validating the condition and triggering a loop. ChatGPT is unable to identify the problem with this code even when it is explicitly prompted in a reply. 

\begin{figure}[t]
\centering
\subfloat[]{%
\resizebox*{8cm}{!}{\begin{tabular}{|c|c|c|}
\hline
  & Small-size solutions & Medium-size solutions \\
  Correct in the first reply & 71\% & 60\% \\
  Correct after follow-up questions & 85\% & 94\% \\ 
  \hline 
\end{tabular}}}\hspace{5pt}
\subfloat[]{%
\resizebox*{7cm}{!}{\includegraphics{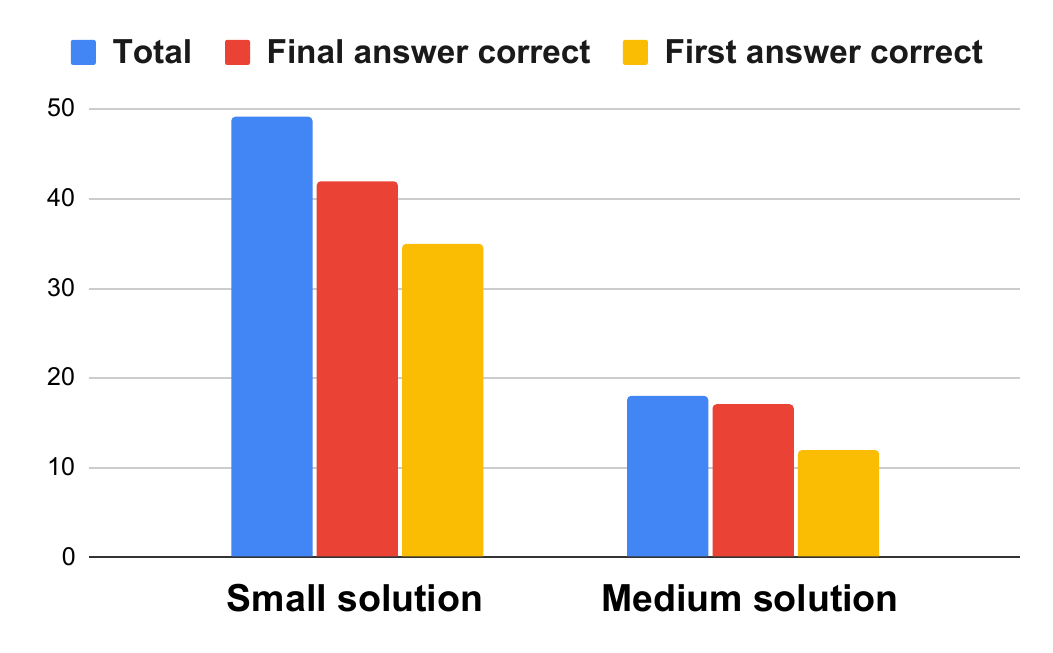}}}
\caption{Correctness of ChatGPT generated solutions per size (lines of code), expressed as: (a) percentages (b) as absolute values from the dataset} \label{size}
\end{figure}

Finally, we looked at the correlation of correctness with the solution size. As before, we observe that there is no sensible correlation here (Figure~\ref{size}). ChatGPT is equally capable in generating small and medium code solutions. Out of 5 exercises with an expected large solution, only 2 received a first-answer solution. After follow-up questions one more correct solution was given. Due to the small number of such exercises, it is hard to draw any conclusions regarding ChatGPT’s ability to generate larger code.
While performing this study we also asked ChatGPT a lot of questions about qualitative aspects of the code it generated. We noticed that it can correctly establish if a given piece of code is written in functional style, if a given function is recursive/tail-recursive or not. This motivated us to experiment with writing a tool which our teaching team could use to semi-automatically generate code reviews over student homework. This tool is described in detail in Section~\ref{sec:tool}.

\subsection{Human vs. code-generated errors}\label{sec:plagiarism}

Our findings suggest that human errors are very different from those found in code generated by ChatGPT. In contrast to ChatGPT, students at the onset of the lecture will have difficulty in writing code that compiles as well as difficulty in solving compilation errors. It is not uncommon for us to observe students bypassing the step of reading and comprehending a compilation error message and, instead, immediately attempting to modify the code. 
The majority of errors that we encounter in compilable code from students stems from an improper functional coding style. For instance, certain students may opt for defining recursive functions to tackle complex tasks that could be more effectively solved by a suitable combination of filtering, mapping and folding (or reducing) operations. These recursive functions can become massive in size (e.g. tens of lines of code) and are often hard to troubleshoot. In such instances we strongly encourage a rewriting of the entire solution.
In contrast, ChatGPT solutions have an overall well-organized functional structure and style, that can be deceptively convincing despite harboring subtle inaccuracies. These errors often originate from the improper utilization of constants, initial values or inadequate consideration of corner cases. As a result, solving such errors may prove challenging, particularly without a robust test-writing methodology in place.

An interesting open question is whether these differences potentially help in distinguishing between student-written code and code that was AI-generated. Although there are discernible error patterns present in both human-generated code and ChatGPT-generated code, we maintain that neither category provides a robust enough signal to facilitate the development of innovative plagiarism detection tools for code generation, as far as our functional programming experience goes.
Changes in style should not be automatically interpreted as instances of plagiarism. Such changes can manifest positively as students progressively develop their functional programming skills over the course of the semester. While they may correlate with some particular instances of plagiarism, they are not proof that unethical coding practice took place. 

We wish to emphasize that plagiarism is an accursation we approach with utmost gravity, reserving it for cases with indisputable evidence. Consequently, a false positive error would be regarded as unacceptable as it could wrongly implicate a student in a plagiarism offense. Conversely, false negatives, while undesirable, are comparatively less severe from an educational perspective as they involve missing instances of plagiarism. At best, tools that build on observations such as ours, which detect changes in coding style, could be used in conjunction with other practices, to ensure ethical conduct in homework submissions.

While code plagiarism detection in the era of AI-generated code is a promising line of research with some new tools available (discussed in Section~\ref{sec:plagiarism-detection}), we currently rely on conventional methods for mitigating such cases, as detailed in Section~\ref{sec:mitigation}.

\section{ChatGPT for code reviews}\label{sec:tool}

To assess the feasibility of using ChatGPT to provide meaningful code reviews to students, we have used student work submitted in the previous academic year. While students were not (yet) given the opportunity to interact with reviews at that point, our results are encouraging and we plan to include such an experiment in the next academic year.

Using our tool, we have queried select pieces of code from 67 homework assignments and asked whether they have been written in functional style. The code pieces ranged in size from 1-2 to a record of 55 lines of code, for the same implementation task. Approx. 77\% of the queries received complete and correct answers, containing valid argumentation. Around 7\% of the answers where overall correct but contained some parts of argumentation that had flaws. For instance, the claim generated by ChatGPT:  ``\emph{the use of return statements to exit early from the function is not idiomatic of functional programming}'' is correct in principle but did not apply to the code piece under scrutiny. Finally, around 15\% of the answers were incorrect.

Our tool was written in Scala. It uses the \cite{ScalaMeta2023} parser to isolate particular pieces of code (e.g. function implementations) from each homework submission. It then assembles the code as well as the question under scrutiny into a ChatGPT query. Finally, it uses the Scala \cite{OpenAIClient2023} which interacts with the OpenAI API using HTTP requests, to submit the query to ChatGPT and retrieve the answer.

\section{Discussion}

\subsection{ChatGPT accuracy} 

Our experiments suggest that ChatGPT is very good at coding in our lab. Although it might seem ideal as a tool for scoring points, ChatGPT is not that good at providing textbook solutions: around half of the correct solutions contain awkward or inefficient pieces of code. While ChatGPTs accuracy is robust against the statement or the solution size, we can see it strongly decrease to around 50\% (a coin toss) for difficult lab assignments, as previously shown. Although some of the existing work on ChatGPT coding accuracy report similar results (see Section~\ref{sec:related}), we should be careful about generalizing these results. FP is a basic programming course, combining learning topics which are widespread on the Internet. For instance, Newton’s Approximation Method is often used as a basic programming exercise. Also, Scala code syntax examples are pervasive online. These would certainly be part of the training corpus for ChatGPT's model. Existing work that has recently emerged on the accuracy of Copilot, a tool that relies on a model sharing similarities with GPT3, suggests that similar performance may be achieved on other, imperative programming languages such as Python.

Based on our findings, we believe that ChatGPT does not currently scale effectively as a learning tool, and relying on its responses as definitive guidance in most cases may do more harm than good. ChatGPT works best when there is a possibility to validate its results e.g. with a credible test suite. However, using ChatGPT to create tests may add to the level of uncertainty regarding its answers.
We do believe that ChatGPT and other open-source LLM models, are a valuable resource to generate code reviews for students as long as it is subject to review itself from human experts. As previously shown, ChatGPT provides ample feedback to our questions, and correctness rates of 77\% indicate that human intervention in writing reviews can be minimized.  

Our datasets for ChatGPT evaluation are publicly available. The link cand be found under references: \cite{ChatGPT2023}.

Finally, we also believe that other LLM-powered tools such as Copilot hold the potential to become promising educational and productivity-enhancing programming aids in the future. We discuss this in more detail in Section~\ref{sec:copilot}.

\subsection{Mitigating ChatGPT in coding assignments}\label{sec:mitigation}
As previously shown, our assessment of ChatGPT's utilization is based exclusively on the responses gathered from students who participated in the survey outlined in Section~\ref{sec:survey}.  At the same time, ensuring good academic ethics is our priority, which requires identifying strategies for mitigating the potential misuse of ChatGPT.

We have not encouraged ChatGPT usage in our lab. In the absence of robust plagiarism tools, our strategy has been to require students to explain, as well as rewrite small, one-line pieces of code under our supervision, in order to acquire points for a homework. In this latter part, we modify one line of code producing a compile error, and ask students to fix it.

This aspect raises interesting guestions, also highlighted by \cite{Fincher2019} in Chapter 14, such as: (i) is the ability of understanding, adapting and altering existing code, whether computer-generated or not, a satisfactory skill for graduating a programming lecture? (ii) from an educational standpoint, is it a beneficial and sustainable practice to learn coding by refining and amending initial code examples, as a complement to the more traditional approach of writing code entirely from scratch?
Although these questions receive diverse responses, each supported with equally compelling arguments, we abstrain from adopting a stance on them, as addresing them lies beyond the scope of this paper.

What we have observed is that students who successfully solved our one-line modification task tend to perform well in our exam and achieve great results in their lab work. It's worth mentioning that during lab sessions, students work on exercises under the guidance of a tutor, hence ChatGPT usage is not likely there. To conclude, our evaluation practice on code ownership aligns quite well with the overall academic student performance in our lecture. Students who have tackled assignments through unethical means typically opt out of presenting their work, resulting in their homework not being graded.

\subsection{GitHub Copilot in coding assignments}\label{sec:copilot}

GitHub Copilot is an LLM-based code generation tool that uses Codex, a model trained on 54 milion software repositories on GitHub and introduced by~\cite{Zaremba2021}. It is integrated as a plugin in code editors such as Visual Studio Code. Although both Copilot and ChatGPT rely on LLMs for their output, they serve different purposes and are even trained on different types of data. Unlike Copilot, ChatGPT was trained on a broad rance of internet text, including websites, public teaching resources, books, articles and more. It is more focused in understanding and generating human language in a general context. Copilot is designed to assist developers in writing code. It provides code suggestions, autocompletions, and documentation as developers write, making it a helpful tool for bug-fixing and software development.
In our survey from Section~\ref{sec:survey} we did not differentiate between the usage of ChatGPT and Copilot. It is evident that Copilot is a more sophisticated tool. Based on our personal experience, at the time when our survey was conducted, most students resorted to ChatGPT in order to get out-of-the-box solutions to coding assignments, which they may adjust on their own subsequently.

In contrast to ChatGPT, Copilot presents itself as a new programming and educational tool with potential, already receiving attention in studies such as those of~\cite{Denny2023}. In the upcoming academic year we plan on considering Copilot as a programming tool in our Functional Programming lecture. Nevertheless, the adoption of such a tool requires a substantial reevaluation of our programming curricula and teaching style. When students rely even more on AI-generated code snippets, new programming skills need to be emphasized. We enumerate a few: breaking a big programming task or algorithm into different parts, better and more in-depth test-writing skills as well as improved code-reading and understanding skills. We believe that this direction is promising an deserves a more comprehensive exploration in future research. This is supported by the usage of Copilot in the programming industry, that has gained momentum. We discuss this in more detail in Section~\ref{sec:DevCopilot}.

\section{Assessment and feedback tools prior to ChatGPT}

Programming has continuously progressed hand in hand with the advancement of its accompanying development tools. From its inception, essential tools like linkers and compilers have played a pivotal role. For example, Grace Hopper's COBOL language evolved as a novel tool allowing programmers to express programms using human-readable instructions rather than laboring with low-level assembly code (\cite{Beyer2009}).

In modern times, programming heavily leans on advanced tools such as Integrated Development Environments (IDEs) which are indispensable for orchestrating complex build systems necessary for large-scale applications. Being aware, accustomed and ultimately profficient in using such tools is an ever important task for the student programmer.

Nodaways, the landscape of educational tools relevant for teaching programming extends far more and includes vizualization tools that are suitable for the illustration of both algorithm and code execution such as that presented by~\cite{Sirkia2009}, learning environments such as Massive Open Online Course platforms (MOOCs), and program development tools which are relevant to both students and senior programs alike (e.g. DrScheme - \cite{Findler1997}).

Reviewing all these approaches is outside the scope of this paper. Instead, we will focus our attention on two types of tools that are pertinent to our work: (i) tools employed by educators to assess aspects such as correctness and coding style in homework submissions and (ii) tools design to assist students in their homework development. These tools can range from straightforward tests made publicly available to students, to more sophisticated testing and grading environments. Sometimes, such tools can have significant overlap with (i).

\subsection{Evaluation and Feedback}\label{sec:eval-feedback}

Computing education research features an extensive body of literature on evaluation and feedback tools, dating back to~\cite{Hollingsworth1960}, where an automatic grader was utilised to evaluate student homework on an IBM 650. Later tools such as ASSESS and AUTOMARK introduced by~\cite{Redish1986} offer more extensive features: ASSESS evaluates specific aspects pertaining to coding style, while AUTOMARK verifies code structure against a template provided by the educator.

Tools such as ASSYST, by~\cite{Jackson1997}, beside correctness achieved by testing, also look at efficiency, style and cyclomatic complexity. At the same time, tools such as PASS, developed by~\cite{Thorburn1997} focus less on testing and more on the approach taken to solve the programming task. Tools such as ASSYST evaluate coding style solely focusing on code indentation and appropriate syntax practices. On the other hand, PASS compares the solution provided by the student with a \emph{solution plan}, in a manner similar to AUTOMARK. Also, the much more recent PASS is designed for C programs, which are notably more intricate compared to the Fortran programs targeted by AUTOMARK.

There is a clear shift from simplistic test-based homework assessment towards a greater emphasis on code quality, encompasing both style and efficiency. This transition is evident in tools such as that of~\cite{Saikkonen2001}, and modern tools such as~\cite{Scalastyle2019}. The work of~\cite{Saikkonen2001}  assesses coding style in homework submissions written in the functional language Scheme, while Scalastyle achieves a similar objective in the functional language of Scala.

\subsection{Homework assistants}

Most, if not all of the previously mentioned tools were primarily designed for use by evaluators. Modern tools used by students (and programmers alike) to aid in the programming task are often provided by IDEs, such as syntax highlighters and syntax completion plugins. In addition to those, our university has a longstanding tradition of offering public test suites for all programming course homework. We follow the same approach in our Functional Programming course. 
The very test suite we employ for grading is also publicly accessible for students, thus ensuring transparency in the grading process. We usually distinguish between two types of tests: (i) those designed to steer students toward the optimal code design solutions. They require defining functions with designated signatures, which are to be reused throughout the homework. Such tests are less significant in the grading process; (ii) tests aimed at verifying functionality of parts of, or the complete homework, typically accounting for the majority of available points in the homework.

This style of test-driven homework development has advantages as well as drawbacks. One significant advantage is that students never have to begin entirely from scratch. The existing code stubs that are part of the test-suite as well as the tests themselves serve as valuable cues on how to commence a homework. On the flip side, a notable disadvantage also pointed out by~\cite{Edwards2003} is that students might become overly reliant on existing tests, potentially hindering their ability to write tests on their own. Another disadvantage that we have often noticed is that students pay less attention to coding style and focus solely on getting points for the majority of tests. This grade-driven approach occasionally yields homework submissions with long, tedious, recursive functions, usually with many parameters, that are hard to follow and are, most of the times, inefficient. Such solutions may receive a passing grade although the written code is inefficient, has little merit with respect  to the functional style that we aim to teach, and usually attest to only a basic familiarity with the syntax of the language.

From an evaluation perspective, applying tools such the ones presented in Section~\ref{sec:eval-feedback} is problematic: most approaches (e.g. ASSYST, PASS) are focused on imperative programming languages; on the other hand, tools for functional programs such as PASS and ScalaStyle focus on syntactic aspects of coding style: they test the usage of certain instructions, which should or should not be used, or the usage of certain programming constructs (e.g. whether or not a class was defined as public, features of a member functions such as being abstract, etc.).

Instead of imposing a given solution template (as for instance, done by PASS), we would like our tool to assess aspects such as: (a) whether certain functions have side-effects or not; (b) whether functional composition or higher-order functions were used in specific parts of the homework; (c) whether some parts of the code can be rewritten using functional composition; (d) if the implementation was suitably broken into smaller parts. From all of the above, only (d) could be properly addressed using existing methods (a variant of cyclomatic complexity could be used for this task).

Furthermore, we would like to make this tool publicly accessible to students, not just evaluators, serving as an assistant to direct students towards a code structure that is efficient, follows functional programming principles and maintains readability.

We anticipate that in the near future, Large Language Models (LLMs) will pave the way for such tools in computing education, in several directions. One of these directions involves generating automated feedback, particularly regarding coding quality. Additionally, tools that seamlessly integrate with popular IDEs like Copilot, aiding students in programming tasks, will significantly influence how both programming and teaching are conducted.

\section{Related work}\label{sec:related}

There is a large public debate covering the future of programming after ChatGPT, such as \cite{Glen2023}. The author argues that, while ChatGPT or Copilot may take over certain repetitive coding tasks, it will not eliminate the need for programmers. Complex programming endeavors such as devising design patterns for programs, creating new algorithms, tackling program computational complexity remain firmly outside the expertise of any generative tool.

\subsection{ChatGPT in education}

Discussions such as the one outlined by~\cite{Lambeets2023}, which revolve around the potential prohibition of ChatGPT in academia, often conclude with a lack of consensus. They produce no concrete outcome especially due to the technical challenges associated with enforcing such a ban.

The study of~\cite{Lau2023} explores the impact of LLMs in education and explores instructors' reactions to issues related to plagiarism using new AI-based tools. According to~\cite{Lau2023}, opinions on their adoption range from advocating for their prohibition to integrating them into educational practices. \cite{Lau2023} provides comprehensive coverage of recent research concerning the utilisation of AI tools in education, including Codex's performance in solving CS1 and CS2 exam problems, the use of AI tools to generate educational content (code explanation, generation and validation of exercises, etc.). This is much in line with our effort for generating code reviews discussed in Section~\ref{sec:tool}. The work of~\cite{Lau2023} goes deeper into how AI tools influenced tutors and instructors and how their personal perspective on these tools will reshape the learning process.

On the other hand, ChatGPTs performance on generating code, especially for the purposes of teaching, has received less attention. The work of \cite{Jalil2023} is one of the few such existing studies. \cite{Jalil2023} is focused on how well ChatGPT performs on both coding questions as well as questions in natural language in the setting of a software testing course. Although it is not directly focused on the code-generation capabilities of ChatGPT, ~\cite{Jalil2023} reports a 53\% accuracy of ChatGPT responses - which is slightly lower that what we have observed. This result is interesting especially since ChatGPT is more adapted to natural language where one would expect to see a better performance.

The work of~\cite{Borji2023} examines the common errors made by ChatGPT when presented with various queries, including those related to programming. Although this study does not primarily deal with evaluating ChatGPTs code generation capabilities, it does shed light on pertinent issues, such at ChatGPT's limitations in tackling intricate programming tasks, its inability to deduce straightforward algorithmic solutions for convoluted programming statements and its inability to deduce mathematical identities. These finding are very much in line with our own observations, as shown in Section~\ref{sec:evaluation}.

\cite{Sobania2023} studies ChatGPT's bug-fixing capabilities with promising results. Their work uses ChatGPT in order to fix bugs in Python programs, as part of the \emph{Automated program repair (APR)} line of research. While their setting is different from ours, a bug-fixing task is very much similar, conceptually, to a programming exercise. \cite{Sobania2023} reports 31 fixed bugs by ChatGPT out of a total of 40, which is a 77.5\% repair rate, quite similar to our reported accuracy.

\subsection{Github Copilot \& Codex}\label{sec:DevCopilot}

Since Copilot's release in October 2021, it has sparked a substantial wave of research examining its impact into the routines and productivity of developers worldwide. This body of research can be broadly categorized into two main groups: (I) studies evaluating the performance of programmers who employ Copilot and (II) studies assessing Copilot's performance across a spectrum of programming tasks.

One of the first such initiatives from the first category belongs to \cite{Vaithilingam2022} whose work is aimed at understanding how programmers use and perceive Copilot. The study compared Intellisense, a VSCode plugin for code completion, with Copilot and revealed that, although Copilot didn't necessarily lead to faster coding completion times, the majority of programmers expressed a preference for integrating Copilot into their daily programming tasks. The primary rationale behind this preference was Copilot's consistent ability to offer valuable starting points and streamline the process by eliminating the need for frequent online searches. At the same time, the study highlighted that participants encountered challenges when it came to comprehending, modifying, and debugging code snippets generated by Copilot. These difficulties had a notable impact on their overall effectiveness in solving tasks. This aligns closely with our observations regarding ChatGPT as well. As previously shown, a substantial portion of the code snippets generated by our model proved to be illegible, particularly for the average student.  A noteworthy finding from this study is the evident over-reliance on Copilot, as exemplified by student P8, who simply accepted the generated code with the remark: \emph{I guess I will take its word}. This further reinforces our perspective that employing generated code without an in-depth understanding of it is not a resilient programming practice. The study belonging to \cite{Vaithilingam2022} was done on 24 participants. 

A more recent study conducted by~\cite{Peng2023} involving 95 participants has yielded a very different outcome. According to their findings, developers who integrated Copilot into their workflow were 55.8\% faster than those who did not utilize it. Notably, developers with limited experience appeared to benefit even more, as they demonstrated quicker task completion times. Their findings emerged within a year of those presented in~\cite{Vaithilingam2022}' study and highlight Copilot's significant and rapid improvement.

Similarly, the work of \cite{Chen2021} is an early evaluation of Codex, for the purposes of Python code generation from documentation, in a  software development setting. \cite{Chen2021} is focused on fine-tuning existing models to achieve better performance, reports similar performance and highlights similar risks such as over-reliance on AI-generated solutions where some code snippets generated by Codex may appear to be correct but are, in fact, flawed.

The study of~\cite{Denny2023} shifted focus from productivity to evaluating Copilot's performance on a publicly accessible dataset comprising 166 programming problems. They found that Copilot successfully solves a around half of these problems on its very first attempt, and further solves the 60\% of the remaining problems solely through natural language adjustments to the problem description. The success rate of Copilot as reported by~\cite{Denny2023}, even when evaluated on imperative programming languages such as Java, Python or C++, closely resembles the results that we have observed. Their work aligns with our obsevations from Section~\ref{sec:copilot}, emphasizing the significance of \emph{prompt engineering}, i.e. the iterative approach of generating and refining code solutions from an initial template, until arriving at a correct solution. This emerging programming methodology holds promise as novel approach to code learning as well as development. However, it requires a more comprehensive evaluation. As both~\cite{Denny2023} as well as our observations suggest, its educational impact deserves more attention, with noteworthy concerns of over-reliance on AI-generated code by beginner programmers.

The paper of~\cite{Kazemitabaar2023} performs a study aimed at assessing how students improve when using Copilot. A total of 69 beginner programmers were tasked with completing 45 coding assignments in Python. Half of the participants used Copilot while the other didn't. The study should an enhancement in the completion rate, with the Copilot users achieving a 1.15x improvements, and an even more substantial 1.8x improvement in their evaluation scores compared to those who did not use Copilot. 

The work conducted by~\cite{Finnie2023} closely aligns with our research goal as it focuses on evaluating Codex's performance in responding to code-writing questions from CS1 and CS2 examinations. The experimental findings presented by~\cite{Finnie2023} remarkably parallel our own in the following respects: (i) Codex's score exhibit consistent decline as the length of generated code increases. In simpler terms, when the solution becomes larger it is more likely to be incorrect. (ii) The Codex performance reported by~\cite{Finnie2023} is around 78\% on the CS1 exam and 58\% in the CS2 exam, with an average of 68\% across exams, which is exactly the same overall success-rate we have observed on ChatGPT. The result is remarkably similar given the differences between experimental settings: the CS1 and CS2 exams are given in Python using imperative and object-oriented coding style and concepts, while our exercises are written in Scala, emphasizing functional, programming principles. Additionally, our exams take the form of lab exercises and a portion of them are larger in both scale and complexity often requiring more time to solve. The similarity of these results to our own suggests that both Codex and GPT 3 models exhibit a robust level of accuracy across beginner and intermediate programming lectures regardless of the programming paradigm.

The work of~\cite{Wermelinger2023} also reports on the usage of Copilot as a possible tool to assist students in solving lab exercises. It confirms previous observations that the first solution generated by Copilot is more likely to be correct. While there is no dataset for accuracy evaluation of Copilot in his work, Wermelinger highlights the abundance of innacurate Copilot solutions. 

\subsection{Generating explanations}

The work of~\cite{MacNeil2023} provides interesting insights similar to our observations from Section~\ref{sec:tool}. Instead of generating code reviews, the work of~\cite{MacNeil2023} is focused on generating explanations for code snippets in programming materials. Three different types of explanations were generated in this study: line-by-line explanations, an enumeration of concepts captured by the code-piece under scrutiny and a general overview of the code's functionality. \cite{MacNeil2023} concludes that explanations were beneficial for students, however, as far as we could observe, there is no reporting of the accuracy or correctness of the LLM-generated explanations.

\subsection{Plagiarism detection}\label{sec:plagiarism-detection}
The work of~\cite{Vahid2023} attempts to build on observations similar to ours (see Section~\ref{sec:plagiarism}), in order to create a plagiarism detection tool for AI generated code. While not focusing on programming errors per se, \cite{Vahid2023} observes that student-written code is very much different in style from that which is AI-generated. In their work, \cite{Vahid2023} tracks each student across  10 different submissions in order to assess changes in coding style. While this approach yields promising results on C/C++ code, as reported by~\cite{Vahid2023}, and it might be a good indicator of some plagiarism instances, we believe it has at least a few limitations: (i) it requires a large number of different submissions (for instance, in our courses, and in most from our University, we have at most 4 project submissions per semester), (ii) it does not take into account the variability of style caused by learning (our students will change coding style over the course of a programming lecture, and we see this as beneficial), (iii) it is inaplicable in instances where solutions are compact: in 3-4 lines of code, there is little formatting, commenting and other syntax-related issues that may be an indicator of a certain coding style. Also, in our experience, students which are aware that their code is checked against plagiarism will adjust their syntax in such a way as to obfuscate (e.g. change variable names, comments, etc.) the real source of their code. One final issue is related to evaluation. The tool described in~\cite{Vahid2023} has been evaluated using a body of students where some admitted to using ChatGPT. It is unclear how an evaluation of such tool could be performed with actual students, as there is, in our view, no 100\% way of knowing for sure that a student has plagiarised using AI tools, unless the admits in doing so.
In our view, while there are approaches which are able to detect AI text generation in natural language, such as those described in~\cite{GPTZero2023}or~\cite{Mitrovic2023}, as far as we are aware, there is still much work to be done towards a robust tool or methodology to determine if a piece of code has been AI-generated. 

\section{Conclusion}

The evolving landscape of research highlights the great potential of LLMs in the realm of programming. It also motivates further exploration to gain more insight into the impact of ChatGPT or Copilot on programming education.

To date, all existing educational research has primarily concentrated on simple, imperative-style programming, predominantly in languages such as Python, occasionally in Java or C/C++. To our knowledge, no investigation except the present one looked at Functional Programming Languages such as Scala, and how these tools perform in such contexts. While our work is an initial step, more are needed to get a comprehensive view of LLMs potential and shortcomings.

Existing plagiarism tools, whether traditional or novel, face limitations in recognising AI-generated code. More importantly in our view, there is scarcity amoung studies exploring how students actually use ChatGPT or Copilot. Also, there exists a great potential in using such tools for generating valuable educational content.

We plan on pursuing the former two directions. More concretely, we plan on organising a new course centered around functional programming that integrates Copilot as a programming tool. Also, we plan on developing an automated process for generating coding style reviews for students, a valuable resource for our future lectures. With these two directions, we aim to lay groundwork for novel approaches in programming education ultimately helping our students become more skilled and productive in the ever-changing landscape of program development.

\section*{Notes on contributors}

\textbf{Matei-Dan Popovici} is an Associate Professor at the National University of Science and Technology POLITEHNICA Bucharest, Romania. He received his PhD degree in Computer Science in 2012 from the same university. His research areas include program and network verification, formal methods, functional programming languages, programming methodologies in education.

\end{document}